\AtBeginDvi{}
\documentclass[onecolumn]{jpsj3}
\usepackage{txfonts}
\usepackage[dvipdfmx]{color}

\title{Electronic Phase Diagram of SrFe$_2$(As$_{1-x}$P$_x$)$_2$: Effect of Structural Dimensionality}

\author{Tatsuya Kobayashi$^1$\thanks{E-mail:kobayashi@tsurugi.phys.sci.osaka-u.ac.jp}, 
Shigeki Miyasaka$^1$, Setsuko Tajima$^1$, and Noriko Chikumoto$^2$}
\inst{$^1$Department of Physics, Graduate School of Science, Osaka University, Toyonaka, Osaka 560-0043, Japan \\
$^2$Superconductivity Research Laboratory-ISTEC, Tokyo 135-0062, Japan \\}

\abst{We have investigated the electronic properties of SrFe$_2$(As$_{1-x}$P$_x$)$_2$ ($0\leqq x\leqq1$) single crystals with postannealing treatment and established a phase diagram of this system. The postannealing treatment reduced the residual resistivity in an antiferromagnetic-orthorhombic phase and enhanced the superconducting transition temperature to $T_{c, optimal}$=33\,K in the superconducting phase. The obtained phase diagram and electronic properties are similar to those of BaFe$_2$(As$_{1-x}$P$_x$)$_2$ despite the large difference in lattice constants. This indicates that, for superconductivity, the dimensionality of the Fermi surface determined by the crystal structural anisotropy is less important than the pnictogen height from the Fe plane.}

\begin{document}
\maketitle

\section{Introduction}
Many types of iron-based superconductors (FeSCs) have been discovered to date, and all of them are layered materials composed of conductive and blocking layers \cite{Stewart}. The electronic band calculations have revealed that a two-dimensional conductive layer results in multiple Fermi surfaces consisting of disconnected cylindrical hole and electron pockets \cite{Hirschfeld}. 
It is theoretically proposed that the nesting between those pockets induces spin-density-wave order in the parent compounds and spin fluctuations that mediate Cooper pairing in the superconducting ones.

AFe$_2$As$_2$ (A=Ba, Sr, Eu, and Ca), one of the parent compounds of FeSC, exhibits the magnetostructural transition from the paramagnetic-tetragonal (PT) phase to the antiferromagnetic-orthorhombic (AFO) phase below the transition temperature $T_\mathrm{S, N}$ \cite{Stewart}. 
Carrier doping by chemical substitution and physical pressure suppress the magnetostructural phase transition and induce superconductivity. 
Another means of inducing superconductivity is the application of chemical pressure by the isovalent substitution of As for P. In particular, BaFe$_2$(As$_{1-x}$P$_x$)$_2$ (P-Ba122) attracts much attention in terms of the nodal superconducting gap and magnetic quantum criticality \cite{Kasahara}. 

According to previous studies of P-A122 (A=Ba, Eu, and Ca) with single crystals \cite{Kasahara, Jeevan, Kasahara2}, as the lattice constants are reduced (Ba$> $Eu$> $Ca), superconductivity is induced with a smaller amount of P-substitution, and the optimal superconducting transition temperature $T_{c, optimal}$ decreases. 
In these systems, it was confirmed that the decrease in the anisotropy ratio of lattice constants ($c/a$) makes the cylindrical Fermi surface more three-dimensional, namely, the electronic anisotropy is reduced \cite{Carrington, Suzuki}. In this work, we use $c/a$ as a measure of the structural and electronic dimensionalities. The observed tendency in P-A122 implies that the anisotropy of the crystal structure and the resultant Fermi surface topology play an essential role in the superconductivity of FeSC. It supports the theoretical model in which the Fermi surface nesting is crucial. 
In contrast to P-A122 (A=Ba, Eu, and Ca), the related compound SrFe$_2$(As$_{1-x}$P$_x$)$_2$ (P-Sr122) has not been studied in detail. 
An early study of polycrystalline P-Sr122 showed the electronic phase diagram up to $x\leqq 0.40$, but the compound with $x\geqq 0.40$ was not obtained \cite{Shi}. 
To clarify the relationship between the crystal structure and the superconductivity in FeSC, it is important to determine whether or not P-Sr122 follows the tendency mentioned above. 

On the other hand, it has been reported that the postannealing treatment sometimes improves the sample quality and induces marked changes in the electronic properties of FeSC \cite{Ishida}. Therefore, the postannealing treatment is necessary to reveal the intrinsic electronic properties of FeSC. 
In a previous study, we found that postannealing treatment reduces the level of disorder, and increases the pnictogen height from the Fe plane in SrFe$_2$(As$_{0.65}$P$_{0.35}$)$_2$, leading to a remarkable enhancement of $T_c$ \cite{Kobayashi2}. 
It is also interesting to investigate the annealing effect on P-Sr122 single crystals over a full range of compositions and to compare their electronic properties with those of other P-A122 systems. 

In this work, we performed a systematic study of the magnetic susceptibility, electric resistivity and Hall resistivity of single crystals of P-Sr122 with $0\leqq x\leqq1$ before and after postannealing treatment. 
We succeeded in growing single crystals of P-Sr122 with $0\leqq x\leqq1$ for the first time and in establishing its complete electronic phase diagram. 
It was found that the postannealing treatment resulted in the reduction in the residual resistivity in the AFO phase and the enhancement of $T_c$ in the superconducting phase, which enlarges the $T_c$ dome in the phase diagram. The optimal $T_c$ reaches 33\,K, which is the highest among those of P-A122 (A=Ba, Sr, Eu, and Ca). The obtained phase diagram is similar to that of P-Ba122 despite the large difference in lattice constants.   

\section{Experimental Procedure}
Single crystals of SrFe$_2$(As$_{1-x}$P$_x$)$_2$ ($0\leqq x\leqq0.6$) were grown by a self-flux method as described in Ref. 10. 
Crystals with a typical size of $1\times 1\times0.1$\,mm$^3$ were obtained. The obtained crystals became smaller as $x$ increased, and eventually became smaller than the measureable size at $x\succ 0.60$. 

For $x=1.0$, single crystals were grown with Sn flux. The molar ratio of the starting materials was 1:2:20 (Sr:FeP:Sn). 
The mixed materials were placed in an alumina crucible and sealed in a silica tube. They were heated to 1300\,$^\circ$C and slowly cooled to 650\,$^\circ$C. 
The flux was decanted in a centrifuge. Crystals of $1\times 1\times0.1$\,mm$^3$ size were obtained. 

The lattice constants of the $a$- and $c$-axes were estimated by single-crystal X-ray diffraction analysis. The compositions of the obtained crystals were determined with an Electron Probe Micro Analyzer (EPMA). The as-grown crystals were sealed in an evacuated silica tube and postannealed under several conditions. The electric resistivity was measured by a standard four-probe method. The measurement of magnetic susceptibility was performed with a superconducting quantum interference device. The Hall resistivity was measured at magnetic fields of up to 7\,T at various temperatures.

\section{Results and Discussion}
  
\begin{figure}
\begin{center}
\includegraphics[width=70mm]{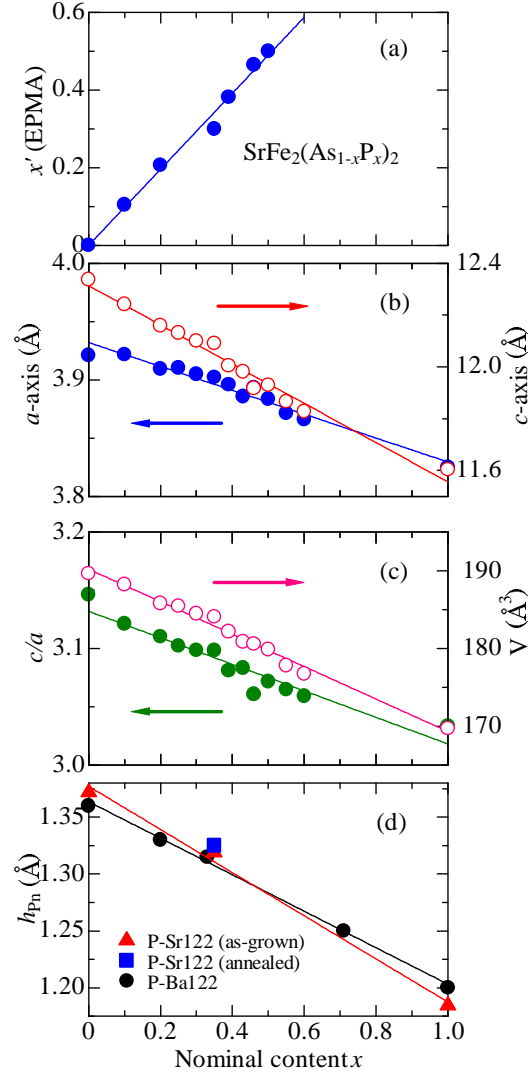}
\end{center}
\caption{(Color online) P-content dependences of (a) actual compositions determined by EPMA in single crystals of SrFe$_2$(As$_{1-x}$P$_x$)$_2$, (b) $a$- and $c$-axes lattice constants, (c) anisotropy ratio $c/a$ and unit cell volume (V), and (d) pnictogen height $h_\mathrm{Pn}$. The data for $x=0$, $1.0$ and P-Ba122 in (d) are cited from Refs. 8 and 3, respectively. The solid lines in (a), (b), (c), and (d) are visual guides.}
\label{Fig1}
\end{figure}

Figure 1(a) shows the P-content determined by EPMA. We use nominal $x$ values hereafter because the actual compositions $x'$ are nearly the same as the nominal ones ($x'=0.979\times x$). 
Figure 1(b) shows the P-content dependences of the lattice parameters for the as-grown crystals. The lattice constants $a$ and $c$ decrease monotonically as the P-content $x$ increases. The anisotropy ratio $c/a$ and the unit cell volume V also decrease with increasing $x$, as shown in Fig. 1(c). These results suggest that P-substitution causes a chemical pressure.

The $c/a$ of SrFe$_2$As$_2$ is largest ($\sim 3.15$) among P-Sr122 but is smaller than that of BaFe$_2$P$_2$, which is the smallest ($\sim 3.24$) among P-Ba122. 
This indicates that the crystal structure of P-Sr122 is less anisotropic than that of P-Ba122. 
Actually, the photoemission measurement revealed that the Fermi surface topology of SrFe$_2$(As$_{0.65}$P$_{0.35}$)$_2$ is more three-dimensional than that of P-Ba122 as well as the crystal structure \cite{Suzuki}. 

Figure 1(d) shows the P-content dependences of the pnictogen height ($h_\mathrm{Pn}$) for P-Sr122 and P-Ba122. Despite the substantial differences in lattice parameters, $h_\mathrm{Pn}$ and its doping dependence are similar in both systems.

\begin{figure}
\begin{center}
\includegraphics[width=70mm]{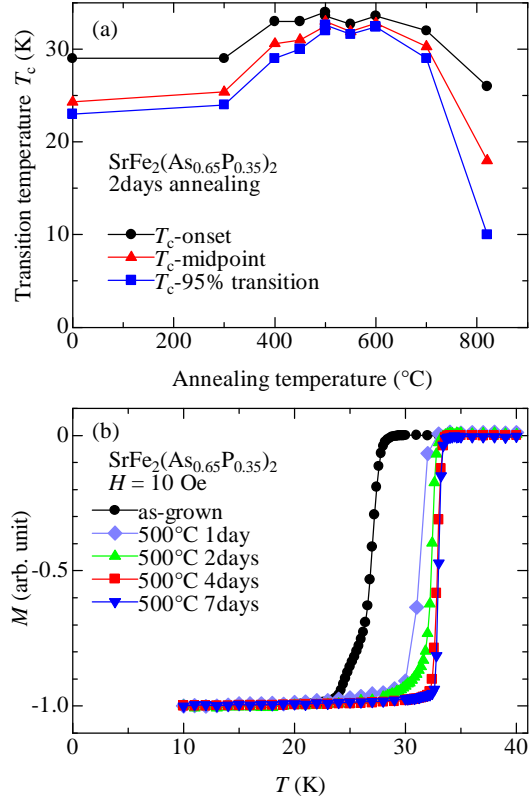}
\end{center}
\caption{(Color online) (a) Annealing temperature dependence of $T_c$ for $x=0.35$. (b) Annealing time dependence of the magnetization $M$ for $x=0.35$ at magnetic field $H=10\,$Oe applied along $c$-axis.}
\label{Fig2}
\end{figure}

To determine the best annealing conditions, we annealed the crystals at various temperatures for various durations. At first, we annealed the crystals of SrFe$_2$(As$_{0.65}$P$_{0.35}$)$_2$ for 2 days at different temperatures. Figure 2(a) shows the annealing temperature dependence of $T_c$ determined from the magnetization $M$. We found that the annealing treatment at 500\,$^\circ$C gives the highest $T_c$ among all the examined temperatures. The annealing treatment at 800\,$^\circ$C damaged the sample and strongly suppressed $T_c$. 
Next, fixing the temperature at 500\,$^\circ$C we annealed the crystals for various durations. The onset $T_c$ reached 33\,K after 4 days of postannealing, and did not change even when they were annealed for longer durations, as shown in Fig. 2(b). Therefore, we decided to anneal all the crystals with $0\leqq x\leqq1$ at 500\,$^\circ$C for 1 week in the present work. 

As we reported previously, we ascribe the $T_c$ increase by our postannealing treatment to the elongation of the pnictogen height \cite{Kobayashi2}. A similar annealing effect was reported in Ca(Fe$_{1-x}$Co$_x$)$_2$As$_2$ \cite{Ran}. Ran $et \,al$. demonstrated that the postannealing treatment increases the $c$-axis lattice constant and its effect depends on the annealing temperature. We speculate that a similar mechanism causes the annealing temperature dependence of $T_c$ in P-Sr122.

 
\begin{figure}
\begin{center}
\includegraphics[width=100mm]{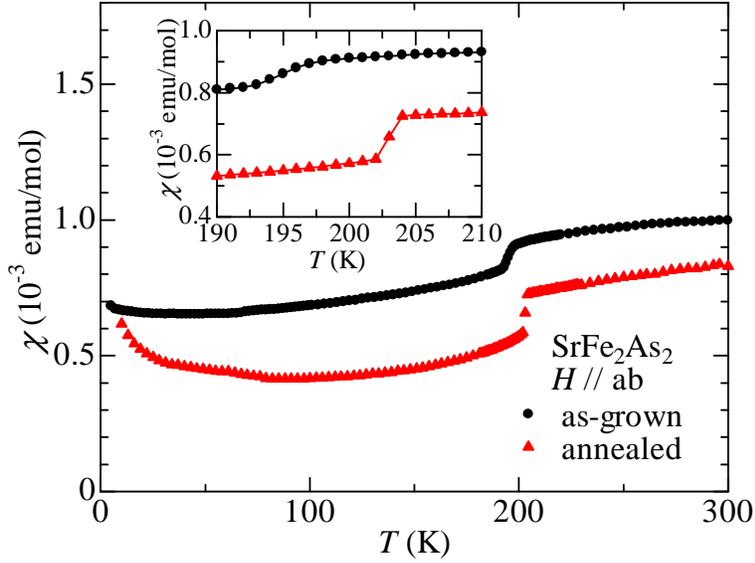}
\end{center}
\caption{(Color online) Temperature dependences of the magnetic susceptibility $\chi$ before and after annealing for $x=0$. The inset shows $\chi$ at approximately $T_\mathrm{S, N}$.}
\label{Fig3}
\end{figure}

Figure 3 shows the temperature dependences of the magnetic susceptibilities $\chi(T)$ of the as-grown and annealed SrFe$_2$As$_2$ samples. The antiferromagnetic transition occurred at $T_\mathrm{S, N}=197$\,K in the as-grown crystal, which is in agreement with a previous report \cite{Yan}. After the annealing treatment, $T_\mathrm{S, N}$ was increased to 204\,K and the transition became sharper, as shown in the inset. The absolute value of $\chi(T)$ in the PT phase was reduced by the postannealing treatment. This behavior was reproducible with different pieces of SrFe$_2$As$_2$. However, it is still unclear what causes the reduction in the magnitude of $\chi(T)$ in the PT phase. The postannealing may be eliminate the paramagnetic impurities or defects and cause the reduction in $\chi(T)$. 
Another possibility is the reduction in the density of state by annealing, which was observed in the electronic specific heat coefficient of the PT phase in SrFe$_2$(As$_{0.65}$P$_{0.35}$)$_2$ \cite{Kobayashi2}, although its mechanism is also unclear.

 
 \begin{figure}
 \begin{center}
\includegraphics[width=100mm]{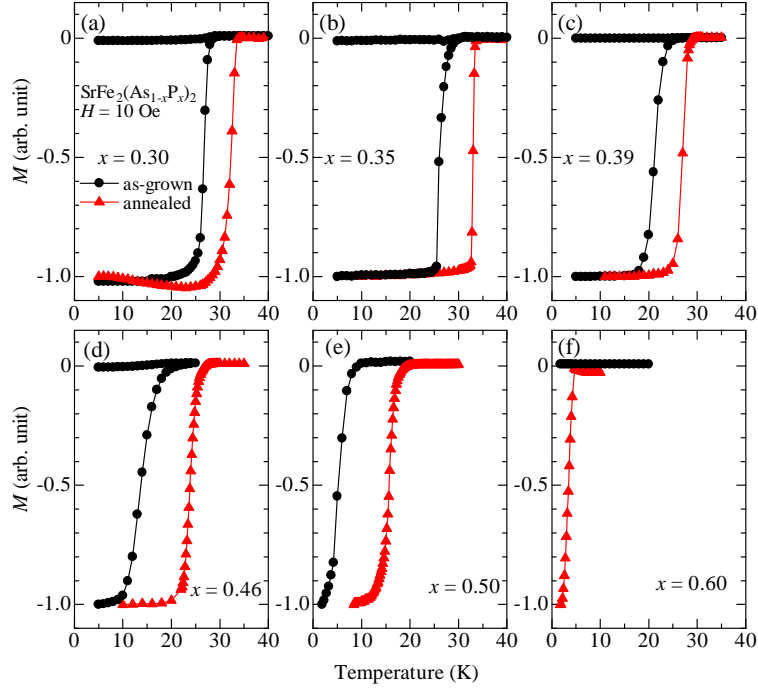}
\end{center}
\caption{(Color online) (a)-(f) Temperature dependence of magnetization $M$ before and after annealing for $x=0.30, 0.35, 0.39, 0.46, 0.50$, and $0.60$ in magnetic field $H$=10\,Oe, respectively.}
\label{Fig4}
\end{figure}
 
The annealing effect on $T_c$ was examined by magnetization $M$ measurement for SrFe$_2$(As$_{1-x}$P$_x$)$_2$ ($0.30\leqq x\leqq0.60$), as shown in Fig. 4. $M$ is normalized by the absolute value at the lowest measured temperature. $T_c$ is clearly increased after the postannealing treatment for all the compositions. For $x=0.30$ and $0.35$, $T_c$ reached 33\,K that is the highest among those of P-A122 (A$=$Ba, Sr, Ca, and Eu) \cite{Kasahara, Nakajima, Jeevan, Kasahara2}. For $x=0.39, 0.46$, and $0.50$, $T_c$ is increased nearly 10\,K after the postannealing. More surprisingly, for $x=0.60$, the annealed crystal showed a superconducting transition at 5\,K, whereas the as-grown one did not exhibit a diamagnetic signal down to 1.8\,K.


\begin{figure}
\begin{center}
\includegraphics[width=100mm]{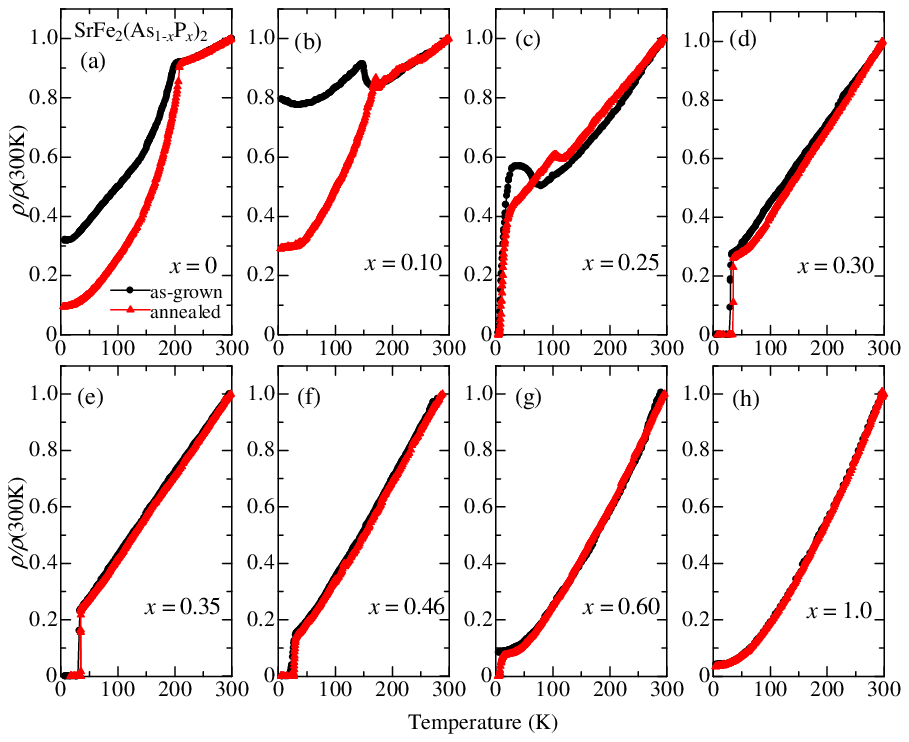}
\end{center}
\caption{(Color online) (a)-(h) Temperature dependence of the resistivity before and after annealing for $0\leqq x\leqq1$. The absolute value of $\rho(300\,\mathrm{K})$ decreases from 0.29\,$\mathrm{m}\Omega\,\mathrm{cm}$ for $x=0$ to 0.22 for $x=0.35$ and 0.030 for $x=1.0$. }
\label{Fig5}
\end{figure}

In Figs. 5(a)-5(h), we show the temperature dependence of resistivity $\rho(T)$ for $0\leqq x\leqq1$. The data are normalized by the value at room temperature, since there was no change in $\rho(300K)$ by the postannealing within a measurement accuracy. 
For $x=0$, although the as-grown crystal showed an anomaly due to the magnetostructural transition at 197\,K, the annealed crystal exhibited a sharper transition at 204\,K, which is consistent with the magnetic susceptibility measurement in Fig. 3. The residual resistivity ratio (RRR) $\rho_\mathrm{300K}$$/\rho_\mathrm{4.2K}$ is also changed from 3 to 10 after the postannealing. The annealing treatment clearly reduced the residual resistivity, indicating that the impurity or disorder within the crystals was removed by this treatment. Also, for $x=0.10$ and 0.25, the residual resistivity clearly decreases and $T_\mathrm{S, N}$ is increased after the annealing treatment [Figs. 5(b) and 5(c)]. For $x\geqq0.35$, at which the magnetostructural transition completely disappears, the reduction in the residual resistivity was smaller than that for $x\leqq0.30$. These indicate that the transport property is more sensitive to the annealing treatment in the AFO phase than in the PT phase. 

For $x=0.35$, the $T$-linear behavior of resistivity is observed, which suggests a non-Fermi liquid state due to a strong spin fluctuation. Actually, at this composition, $(T_1T)^{-1}$ is strongly enhanced at low temperatures, where $T_1$ is the spin-lattice relaxation rate in nuclear magnetic resonance (NMR) \cite{Duluguun}. With further P doping, $\rho(T)$ changed to the Fermi-liquid-like $T^2$ dependence. Moreover, the enhancement of $(T_1T)^{-1}$ was not observed at $x=0.50$. These results suggest that a two-dimensional antiferromagnetic quantum critical point exists at approximately $x=0.35$ in P-Sr122. Similar quantum critical behaviors were also reported in the P-Ba122 \cite{Kasahara} and P-Eu122 \cite{Jiang}, indicating that such a quantum criticality is a common feature in the P-doped 122 system. 


\begin{figure}
\begin{center}
\includegraphics[width=80mm]{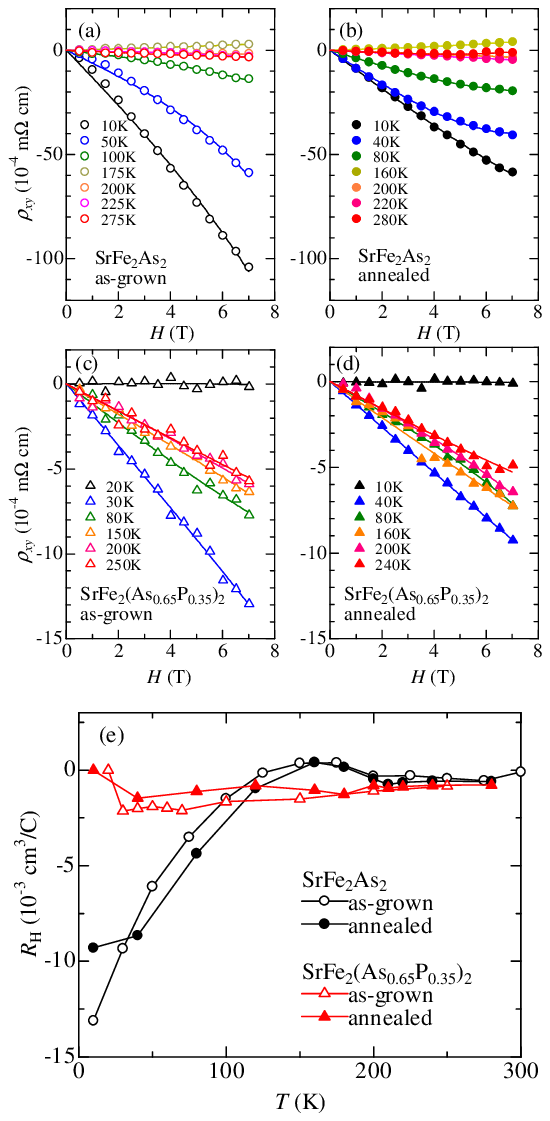}
\end{center}
\caption{(Color online) (a)-(d) Magnetic field ($H$) dependence of the Hall resistivities ($\rho_{xy}$) of as-grown and annealed single crystals for $x=0$ and 0.35, respectively. The solid lines indicate the fitting results by the formula $\rho_{xy}=R_\mathrm{H}H+aH^3$. (e) Temperature dependence of Hall coefficient $R_\mathrm{H}$ of as-grown (open symbols) and annealed (closed ones) single crystals for $x=0$ (circles) and 0.35 (triangles), respectively. The solid lines are visual guides.}
\label{Fig6}
\end{figure}

To obtain further information on the electronic state, we studied the Hall effect in P-Sr122 with $x=0$ and $0.35$. Figure 6 shows the magnetic field dependence of the Hall resistivity $\rho_{xy}(H)$ and the temperature dependence of the Hall coefficient $R_\mathrm{H}(T)$ for $x=0$ and 0.35. $R_\mathrm{H}(T)$ was estimated by polynomial fitting with $\rho_{xy}=R_\mathrm{H}H+\alpha{H}^3$. For $x=0$ [Figs. 6(a) and 6(b)], the absolute value of $\rho_{xy}(H)$ decreases and deviates from the $H$-linear dependence below $T_{S, N}$ after the annealing treatment. A similar result was reported in a previous study of the annealed BaFe$_2$As$_2$ \cite{Ishida}. However, the $H$ dependence of $\rho_{xy}$ is weaker in the annealed SrFe$_2$As$_2$ than in the annealed BaFe$_2$As$_2$. This may be caused by the smaller RRR of the annealed SrFe$_2$As$_2$ (RRR$\sim$10) than of the annealed BaFe$_2$As$_2$ (RRR$\sim30$) \cite{Ishida}. In Figs. 6(c) and 6(d) for $x=0.35$, $\rho_{xy}(H)$ shows an $H$-linear dependence and little change with postannealing, which is consistent with the annealing effect on $\rho(T)$ shown in Fig. 5. 

In Fig. 6(e), the estimated $R_\mathrm{H}(T)$ value for $x=0$ and 0.35 are plotted. For SrFe$_2$As$_2$, $R_\mathrm{H}(300\,\mathrm{K})$ is negative, indicating that the dominant carriers are electrons. At approximately $T_\mathrm{S, N} (\sim200\,\mathrm{K})$, $R_\mathrm{H}(T)$ changes its sign. 
Such a sign change of $R_\mathrm{H}$ is absent in BaFe$_2$As$_2$ where $R_\mathrm{H}(T)$ abruptly decreases at $T_{S, N}$ \cite{Fang}, while it was present in EuFe$_2$As$_2$ and CaFe$_2$As$_2$ \cite{Matusiak, Matusiak2}. Such a behavior may be ascribed to the presence of the Dirac fermion \cite{Morinari}, which was observed in SrFe$_2$As$_2$ by quantum oscillation measurement \cite{Sutherland}. 
$R_\mathrm{H}(T)$ markedly decreases below $T_{S, N}$, and the absolute value of $R_\mathrm{H}(T)$ at the lowest temperature is 24 times larger than that at room temperature, indicating that the carrier reduction is caused by the reconstruction of the Fermi surfaces in the AFO phase. Although the carrier concentration is reduced, the resistivity shows a sharp decreases at $T_{S, N}$ and maintains a metallic behavior, as shown in Fig. 5. This reflects the multicarrier nature of iron pnictides. 
For $x=0.35$, although the magnetostructural transition completely disappears, the absolute value of $R_\mathrm{H}(T)$ is slightly enhanced by a factor of 2 as the temperature decreases. 
This may be caused by the multiband effect \cite{Fang} or the vertex correction due to antiferromagnetic fluctuation \cite{Fanfarillo}.


\begin{figure}
\begin{center}
\includegraphics[width=120mm]{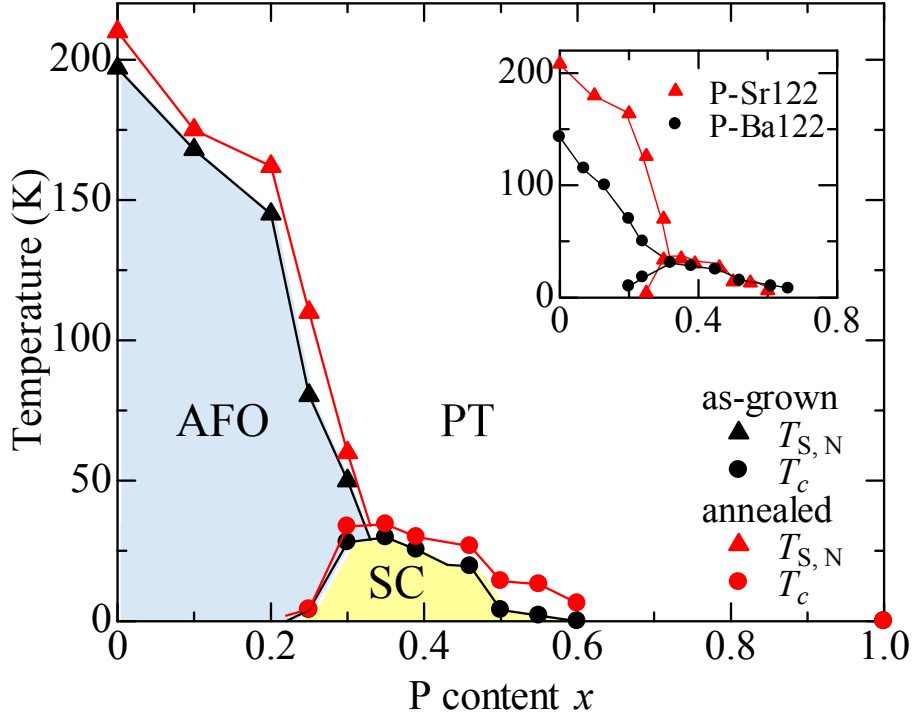}
\end{center}
\caption{(Color online) Phase diagram obtained from the as-grown and annealed single crystals of SrFe$_2$(As$_{1-x}$P$_x$)$_2$. $T_\mathrm{S, N}$ is defined as the temperature at which resistivity shows an anomaly due to the phase transition to the AFO state. $T_c$ is defined as the temperature at zero resistivity. The inset compares the phase diagrams of annealed P-Sr122 and P-Ba122 systems. The data of P-Ba122 is cited from Ref. 13.}
\label{Fig7}
\end{figure}
 
The electronic phase diagram obtained from the susceptibility and resistivity measurements is shown in Fig. 7. Roughly speaking, this is the same as the general phase diagram observed in AFe$_2$As$_2$ compounds. Namely, with P-substitution, the magnetostructural transition is gradually suppressed and completely disappears at $x=0.35$, while the superconductivity emerges above $x=0.25$. For the as-grown crystals, $T_{c, optimal}=26\,\mathrm{K}$ and the superconductivity disappears above $x=0.60$. When the crystals are postannealed, $T_\mathrm{S, N}$ increases and the superconducting dome is enlarged.
For $x=0.25$ and 0.30, both $T_\mathrm{S,N}$ and $T_c$ are increased by the annealing treatment. For $x=0.35$, the optimal $T_c$ is enhanced to 33\,K by postannealing. The obtained optimal $T_c$ is as high as that of the physical-pressure-induced superconductivity in SrFe$_2$As$_2$ and higher than that (30\,K) of BaFe$_2$As$_2$ \cite{Kotegawa, Colombier}. This means that the chemically or physically pressured Sr122 has a higher optimal $T_c$ than Ba122. In the annealed sample with $x=0.60$, both zero resistivity and a diamagnetic signal were observed at approximately 5\,K, while the as-grown crystal showed no superconductivity signature down to 1.8\,K. The results indicate that the superconducting region is widened beyond $x=0.60$ by the annealing treatment.

The phase diagram for the as-grown crystals shows a smaller superconducting dome than that of P-Ba122. This seems to be ascribed to the more three-dimensional structure and the Fermi surface topology in P-Sr122 than in P-Ba122. However, after the postannealing treatment, the obtained $T_c$ dome is enlarged and very similar to that of P-Ba122, despite the large difference in the $a$- and $c$-axes lattice constants. This is in contrast to the case of P-A122 (A=Eu and Ca).
The recent photoemission measurement has revealed that the hole Fermi surface with the $d_\mathrm{z^2}$ orbital character is strongly warped along the $k_\mathrm{z}$-direction in P-Sr122 \cite{Suzuki}. The more three-dimensional electronic state realized in P-Sr122 should result in the worse Fermi surface nesting than that in P-Ba122, and thus a lower $T_c$ if the nesting condition is crucial for superconductivity. Nevertheless, the optimal $T_c$ of P-Sr122 is nearly the same or even slightly higher than that of P-Ba122. This indicates that $T_c$ is rather insensitive to the crystal structural anisotropy $c/a$ and the dimensionality of the electronic structure.

In contrast to the large difference in the lattice constants, the pnictogen heights are nearly the same in the optimal doped P-Sr122 and P-Ba122 ($x\sim0.3$), $h_\mathrm{Pn}\sim1.32$\,\AA \cite{Kobayashi2, Kasahara}, as shown in Fig. 1(d). 
In FeSC, it is well known that the pnictogen height and As-Fe-As bond angle correlate with $T_c$ \cite{Lee, Mizuguchi}. 
In our previous work, we found that the postannealed P-Sr122 has a longer pnictogen height and cleaner electronic state than the as-grown one \cite{Kobayashi2}.  
The close relationship between the superconductivity and the pnictogen height has been explained by the theory based on antiferromagnetic fluctuation mechanism \cite{Kuroki}. This theory predicts that the reduction in the $a$- and $c$-axes lattice constants generally decreases $T_c$, while the elongation of $h_\mathrm{Pn}$ enhances it. The present results demonstrate that the latter factor ($h_\mathrm{Pn}$) is more important than the former ($a$ and $c$) for superconductivity, which gives some constraints on the theory of superconductivity in FeSC. 

\section{Conclusions}
 We have successfully grown single crystals of SrFe$_2$(As$_{1-x}$P$_x$)$_2$ over the entire range of $x$ and performed a systematic study of its electronic properties as well as of the annealing effects. The annealing treatment reduced the residual resistivity and increased $T_{S, N}$ and $T_c$. Using the annealed crystals, we have established the intrinsic phase diagram that was quite similar to that of P-Ba122 despite the large difference in the $a$- and $c$-axes lattice constants. The similarity between the two systems might be attributed to the nearly identical pnictogen height, which dominates the electronic properties of iron pnictides. The dimensionality of the Fermi surface is a less important factor for determining $T_c$. 

\begin{acknowledgment}
The present work was supported by JST TRIP and IRON-SEA. T. K. acknowledges the Grant-in-Aid for JSPS Fellows.



\end{acknowledgment}



\begin{thebibliography}{99}
 \bibitem{Stewart} G. R. Stewart, Rev. Mod. Phys. \textbf{83}, 1589 (2011).
   \bibitem{Hirschfeld} P. J. Hirschfeld, M. M. Korshunov, and I. I. Mazin, Rep. Prog. Phys. \textbf{74}, 124508 (2011).
\bibitem{Kasahara} S. Kasahara, T. Shibauchi, K. Hashimoto, K. Ikada, S. Tonegawa, R. Okazaki, H. Shishido, H. Ikeda, H. Takeya, K. Hirata, T. Terashima, and Y. Matsuda, Phys. Rev. B \textbf{81}, 184519 (2010).
 \bibitem{Jeevan} H. S. Jeevan, D. Kasinathan, H. Rosner, and P. Gegenwart, Phys. Rev. B \textbf{83}, 054511 (2011).
\bibitem{Kasahara2} S. Kasahara, T. Shibauchi, K. Hashimoto, Y. Nakai, H. Ikeda, T. Terashima, and Y. Matsuda, Phys. Rev. B \textbf{83}, 060505(R) (2011).
\bibitem{Carrington} A. Carrington, Rep. Prog. Phys. \textbf{74}, 124507 (2011).
\bibitem{Suzuki} H. Suzuki, T. Kobayashi, S. Miyasaka, T. Yoshida, K. Okazaki, L. C. C. Ambolode II, S. Ideta, M. Yi, M. Hashimoto, D. H. Lu, Z.-X. Shen, K. Ono, H. Kumigashira, S. Tajima, and A. Fujimori, Phys. Rev. B \textbf{89}, 184513 (2014).
 \bibitem{Shi} H. L. Shi, H. X. Yang, H. F. Tian, J. B. Lu, Z. W. Wang, Y. B. Qin, Y. J. Song and J. Q. Li, J. Phys.: Condens. Matter \textbf{22}, 127502 (2010).

\bibitem{Ishida} S. Ishida, T. Liang, M. Nakajima, K. Kihou, C. H. Lee, A. Iyo, H. Eisaki, T. Kakeshita, T. Kida, M. Hagiwara, Y. Tomioka, T. Ito, and S. Uchida, Phys. Rev. B \textbf{84}, 184514 (2011).
 \bibitem{Kobayashi2} T. Kobayashi, S. Miyasaka, S. Tajima, T. Nakano, Y. Nozue, N. Chikumoto, H. Nakao, R. Kumai, and Y. Murakami, Phys. Rev. B \textbf{87}, 174520 (2013).
 \bibitem{Ran} S. Ran, S. L. Bud'ko, W. E. Straszheim, J. Soh, M. G. Kim, A. Kreyssig, A. I. Goldman, and P. C. Canfield, Phys. Rev. B \textbf{85}, 224528 (2012).
\bibitem{Yan} J.-Q. Yan, A. Kreyssig, S. Nandi, N. Ni, S. L. Bud'ko, A. Kracher, R. J. McQueeney, R. W. McCallum, T. A. Lograsso, A. I. Goldman, and P. C. Canfield, Phys. Rev. B \textbf{78}, 024516 (2008).
 \bibitem{Nakajima} M. Nakajima, S. Uchida, K. Kihou, C. H. Lee, A. Iyo, and H. Eisaki, J. Phys. Soc. Jpn. \textbf{81}, 104710 (2012).
\bibitem{Duluguun} T. Dulguun, H. Mukuda, T. Kobayashi, F. Engetsu, H. Kinouchi, M. Yashima, Y. Kitaoka,
S. Miyasaka, and S. Tajima, Phys. Rev. B \textbf{85}, 144515 (2012).
 \bibitem{Jiang} S. Jiang, H. S. Jeevan, J. Dong, and P. Gegenwart, Phys. Rev. Lett. \textbf{110}, 067001 (2013).
\bibitem{Fang} L. Fang, H. Luo, P. Cheng, Z. Wang, Y. Jia, G. Mu, B. Shen, I. I. Mazin, L. Shan, C. Ren,
 and H. H. Wen, Phys. Rev. B \textbf{80}, 140508(R) (2009).
 \bibitem{Matusiak} M. Matusiak, Z. Bukowski, and J. Karpinski, Phys. Rev. B \textbf{83}, 224505 (2011). 
\bibitem{Matusiak2} M. Matusiak, Z. Bukowski, and J. Karpinski, Phys. Rev. B \textbf{81}, 020510 (2010).
 \bibitem{Morinari} T. Morinari, E. Kaneshita, and T. Tohyama, Phys. Rev. Lett. \textbf{105}, 037203 (2010).
\bibitem{Sutherland} M. Sutherland, D. J. Hills, B. S. Tan, M. M. Altarawneh, N. Harrison, J. Gillett, E. C. T. O'Farrell, T. M. Benseman, I. Kokanovic, P. Syers, J. R. Cooper, and S. E. Sebastian, Phys. Rev. B \textbf{84}, 180506(R) (2011).
 \bibitem{Fanfarillo} L. Fanfarillo, E. Cappelluti, C. Castellani, and L. Benfatto, Phys. Rev. Lett. \textbf{109}, 096402 (2012).
\bibitem{Kotegawa} H. Kotegawa, H. Sugawara, and H. Tou, J. Phys. Soc. Jpn. \textbf{78}, 013709 (2009).
 \bibitem{Colombier} E. Colombier, S. L. Bud'ko, N. Ni, and P. C. Canfield, Phys. Rev. B \textbf{79}, 224518 (2009).
 \bibitem{Lee} C. H. Lee, A. Iyo, H. Eisaki, H. Kito, M. T. Fernandez-Diaz, T. Ito,
K. Kihou, H. Matsuhata, M. Braden, and K. Yamada, J. Phys. Soc. Jpn. \textbf{77}, 083704 (2008).
\bibitem{Mizuguchi} Y. Mizuguchi, Y. Hara, K. Deguchi, S. Tsuda, T. Yamaguchi, K. Takeda, H. Kotegawa, H. Tou, and Y. Takano, Supercond. Sci. Technol. \textbf{23}, 054013 (2008).
\bibitem{Kuroki} K. Kuroki, H. Usui, S. Onari, R. Arita, and H. Aoki, Phys. Rev. B \textbf{79}, 224511 (2009).

\end{thebibliography}
\end{document}